\documentclass[prd,nofootinbib,tightenlines,showkeys,showpacs]{revtex4}
\usepackage{amsfonts}
\usepackage{amssymb}
\usepackage{amsmath}
\usepackage{indentfirst}
\usepackage[latin1]{inputenc}

\begin{document}
\title{Maxwell's equations on the $\kappa$-Minkowski spacetime and Electric-Magnetic duality }

\author{E. Harikumar \footnote{harisp@uohyd.ernet.in}}
\affiliation{School of Physics, University of Hyderabad, Central University P O, Hyderabad, AP, India, PIN-500046}
\vspace*{1cm}
\begin{abstract}
We derive the Maxwell's equations on the $\kappa$-deformed spacetime, valid up to first order in the deformation parameter, using the Feynman's approach.  We show that the electric-magnetic duality is  a symmetry of these equations. It is also shown that the laws of electrodynamics are {\it different} for particles of equal charges, but with different masses. We show that the Poincare angular momentum, required to maintain the usual Lorentz algebra structure, do not get any $\kappa$-dependent corrections.
\end{abstract}


\maketitle

\section{ Introduction}
Noncommutative spacetime was originally introduced to cure the IR divergences that plague quantum theories\cite{snyder}. Though this approach was not pursued further due to the spectacular success of renormalisation programme, Connes's work has rekindled the interest in noncommutative (NC) spacetime and physics on such manifolds\cite{connes}.  It was shown that noncommutative gauge theory defined on Moyal spacetime naturally emerges as a limit of certain string theory model. A mapping between the NC gauge theory and a gauge theory defined on the commutative spacetime was also obtained \cite{sw} and this led to intense activities in construction and study of various physical models on NC spacetime. The field theory models on NC spacetime have highly non-local and non-linear interactions which are not present in the corresponding commutative theories. UV/IR mixing, which is a characteristic feature of field theories on  NC spacetime raises serious questions about the renormalisability of such theories and many authors have investigated  these aspects \cite{renor}. Noncommutative spacetime  also provides a way to model the spacetime uncertainties  which arises due to the quantum gravity effects\cite{sergiod}. 

The  widely studied and by now well understood example of noncummutative spacetime is the Moyal spacetime, whose coordinates obey $[{\hat X}^\mu, {\hat X}^\nu]=i\theta^{\mu\nu}$, where $\theta^{\mu\nu}$ is a constant. One of the most striking features of the NC spacetime brought out in recent times is the fact that the symmetries of these models are realised in terms of Hopf algebra \cite{chaichian}. NC spacetime with coordinates satisfying a Lie algebra type commutation relation is another class of spacetimes that is being investigated vigorously. An example for a spacetime of this class is the fuzzy sphere\cite{madore} which is being studied with various motivations\cite{balbook}. Another prototype of this class is the $\kappa$-deformed spacetime whose coordinates obey $[{\hat x}^i, {\hat x}^j]=0, [{\hat x}^0,{\hat x}^i]=a x^i , (a=\frac{1}{\kappa})$\cite{lukierski}. $\kappa$-spacetime and its symmetry algebra naturally appears in the context of doubly special relativity (DSR) \cite{dsr} (a modified relativity principle having one more dimensionfull parameter  other than the velocity of light). $\kappa$-Poincare algebra and DSR are also known to be related to certain quantum gravity models\cite{dsr}. This led to investigations on $\kappa$-spacetime and physics on $\kappa$-spacetime in recent times\cite{wess,sm,us}, bringing out many interesting aspects of $\kappa$-spacetime and physics models on $\kappa$-spacetime. 

Recently Klein-Gordon equation have been derived on $\kappa$-spacetime  and various aspects of this model have been analysed\cite{lukierski,wess,sm}. It is known that there are many different proposals for Klein-Gordon equation in $\kappa$-spacetime \cite{lukierski, wess,sm}, all satisfying the criterion of invariance under $\kappa$-Poincare algebra. There are some works where $U(1)$ gauge theory on 
$\kappa$-spacetime had been constructed\cite{mdljlm}. The $U(1)$ theory in $\kappa$-spacetime is constructed in terms of commutative fields using $*-$product and Seiberg-Witten map, up to first order in the deformation parameter.  

Here we take a different approach-Feyman's approach, to derive the Maxwell's equations on $\kappa$-spacetime.  In this framework, starting with Newtons second law of motion and commutation relations between coordinates and velocities, Feynman  obtained homogeneous Maxwell's equations \cite{fd}. Various aspects of this approach have been investigated, including the generalisations to obtain non-abelian gauge theories, in depth \cite{fap}. Tanimura had obtained relativistic covariant generalisation of this approach\cite{tanimura}. It was also shown that the only possible interactions a quantum mechanical particle can have are those with scalar, gauge and gravitational fields\cite{tanimura}. In recent times, the Feynman's approach has been generalised to obtain inhomogeneous Maxwell's equations, in addition to the homogeneous ones.\cite{jmp40}. The inherent non locality results in novel interactions in non-commutative theories (which are not shared by their commutative counterparts).  In \cite{plaijtpa}, it was shown that the $so(3)$ symmetry is broken in the Feynman approach because of the non vanishing commutators between the velocities and coordinates. But this symmetry can be restored by augmenting the generators by magnetic angular momentum. Thus it is of interest to apply Feynman's approach to non-commutative spacetime and derive the corresponding Maxwell's equations as well as to see the allowed interactions. The Feynman's approach we adopt here has been used to obtain Maxwell's equation in Moyal spacetime also for more general cases \cite{cf}. 

In this paper, we derive both homogeneous and inhomogeneous Maxwell's equations in $\kappa$-spacetime, using the covariant approach of \cite{tanimura}, adapted to noncommutative spacetime. We obtain the $\kappa$ dependent correction to Maxwell's equations up  to {\it first order} in the deformation parameter as well as the modified force equation. From the structure of the force equation, we argue that the interactions of the particle with scalar and gauge fields do get $\kappa$ dependent modifications. Using explicit form of the $\kappa$-deformed Maxwell's equation, we then show that the $\kappa$-Maxwell's equations are invariant under  the electric-magnetic duality transformation, as in the commutative spacetime. We also obtain the generators of the undeformed Lorentz algebra and show that this naturally leads to the introduction of magnetic monopole.

This paper is organised as follows. After a brief summary of the Feynman's approach in the covariant form in section 2, we present the derivation of Maxwell's equation in $\kappa$-spacetime, valid up to first order in the deformation parameter, in section 3. We also derive and analyse the force equation here. In section 4, we discuss the electric-magnetic duality of $\kappa$-deformed Maxwell's equation (up to first order in the deformation parameter). We then discuss the restoration of the Lorentz algebra in section 5. Our concluding remarks are given in section 6. 

\section{ Feynman's approach to Maxwell's equations}
In this section, we present a short summary of essential details of the covariant generalisation of Feynman's approach\cite{tanimura}. This approach makes certain basic assumption \cite{tanimura} in deriving Maxwell's equations. One is that the coordinates of the relativistic particle in $4$-dimensions is described by $x^\mu(\tau)$, where $\tau$ is just a parameter and not the proper time.The coordinates have vanishing commutation relations between them,i.e.,
\begin{equation}
[x^\mu, x^\nu]=0.\label{com1}
\end{equation}
Another important assumption is the Feynman bracket between the coordinates and the velocities, i.e.,
\begin{equation}
[x^\mu, {\dot x}^\nu]=\frac{i\hbar}{m}\eta^{\mu\nu}\label{com2}
\end{equation}
where ${\dot x}^\mu=\frac{d x^\mu}{d\tau}$. The above equation correctly reproduces the familiar commutation relation between the coordinates and momenta.  But we note that in original approach of Feynman's \cite{fd}, existence of momentum was not assumed.  

Thirdly, it is assumed that the Newton's equation holds good, i.e.,
\begin{equation}
m{\ddot x}^\mu=F^\mu(x,{\dot x}),\label{NE}
\end{equation}
where $F^\mu$ can depend on velocity also. Last assumption is that the brackets involving coordinates and velocities satisfy
\begin{equation}
(i)~~~~~ [A,B]=-[B,A]~~~~~{(\rm anti-symmetry)}\label{asym},
\end{equation}
\begin{equation}
(ii)~~~[A ,[B,C]]+[B,[C,A]]+[C,[A,B]]=0~~~~{(\rm Jacobi~ Identity)}\label{jacobi},
\end{equation}
and the Leibniz rules,
\begin{eqnarray}
(iii)~~~ [A, BC]&=&[A, B]C+B[A, C],\nonumber\\
(iv)~~~ \frac{d}{dt}[A, B]&=&[\frac{A}{dt}, B]+[A, \frac{dB}{dt}].\label{leib}
\end{eqnarray}
It is known that last property listed above is not automatically satisfied by Poisson brackets, unless canonical equations of motion are assumed\cite{cf}. This is the reason for using commutators rather than Poisson brackets in this approach.

Combining Eqns.(\ref{com1},\ref{com2}) with Leibniz rules in Eqn.(\ref{leib}) leads to useful identities 
\begin{eqnarray}
&[x^\mu, f(x^\nu)]=0,~~~~~[x^\mu, f(x^\nu, {\dot x}^\lambda)]=\frac{i\hbar}{m}\frac{\partial f(x^\mu, {\dot x}^\nu)}{\partial{\dot x}^\nu}&\\
&[{\dot x}^\mu, f(x^\nu, {\dot x}^\lambda)]=-\frac{i\hbar}{m}\frac{\partial f}{\partial x^\mu}.\label{iden}&
\end{eqnarray}

A second rank anti-symmetric tensor is naturally introduced by considering the differentiation of Eqn.(\ref{com2}), Thus we find,
\begin{equation}
[{\dot x}^\mu, {\dot x}^\nu]=-[x^\mu, {\ddot x}^\nu]=\frac{iq\hbar}{m^2} F^{\mu\nu}.\label{fs}
\end{equation}
This $F^{\mu\nu}$ would be identified with the Maxwell field strength and $q$ with the electric charge of the particle. 

Now repeated use of the Jacobi identities involving coordinates, velocities as well as those involving both along with judicious use of Eqn.(\ref{iden}) leads to the homogeneous Maxwell's equations
\begin{equation}
\partial^\mu F^{\nu\lambda}+\partial^\nu F^{\lambda\mu}+\partial^\lambda F^{\nu\mu}=0\label{homome}
\end{equation}
where $F^{\mu\nu}=\partial^\mu A^\nu(x)-\partial^\nu A^\mu(x).$ 

It is also shown that the force $F^\mu(x,{\dot x})$ appearing in Eqn.(\ref{NE}) above can be expressed in terms of $F^{\mu\nu}$ and a vector quantity as
\begin{equation}
F^\mu(x,{\dot x})=G^\mu + qF^{\mu\nu}{\dot x}_\nu\label{force}
\end{equation}
where $G^\mu$ is the gradient of a scalar field, i.e., $G^\mu=\partial^\mu \Phi$ and $q$ is the electric charge of the particle.

 Thus, in this approach, one gets the homogeneous Maxwell's equations and Lorentz force equation, starting with the assumption of the Feynman bracket between coordinates and velocities.

This construction was generalised to curved spacetime by modifying Eqn.(\ref{com2}) by replacing $\eta^{\mu\nu}$ with $g^{\mu\nu}$. In this case, it was shown that the force $F^\mu$ in Eqn.(\ref{force}) gets an additional contribution which depends on the derivative of metric tensor $g^{\mu\nu}$\cite{tanimura}. Thus, in this approach, two of the Maxwell's equations were obtained in a manifestly covariant form. This also showed that the particles get coupled naturally to a scalar, gauge, and gravitational fields. Later, in \cite{jmp40}, it was shown that the inhomogeneous Maxwell's equations can also be derived from Eqn.(\ref{homome}) itself by contracting any two of the three indices. Thus contracting $\mu$ with $\nu$ in Eqn.(\ref{homome}) sets
\begin{equation}
\partial_\mu F^{\mu\nu}=j^\nu
\end{equation}
where $j^\nu$ is the conserved current, a fact which can be seen by using the Jacobi identity given in Eqn.(\ref{jacobi}) and Eqn.(\ref{iden}).

\section{$\kappa$-deformed Maxwell's equations}
In this section, we apply the covariant generalisation of Feynman's approach of above section to $\kappa$-deformed spacetime whose coordinates obey
\begin{equation}
[{\hat x}^\mu, {\hat x}^\nu]=i c^{\mu\nu}_{~~\lambda}{\hat x}^\lambda\label{kappacoord}
\end{equation}
where
\begin{equation}
c^{\mu\nu}_{~~\lambda}{\hat x}^\lambda=(a^\mu\eta^{\nu}_{~\lambda}-a^\nu\eta^{\mu}_{~\lambda})x^\lambda, a^0=a={\kappa}^{-1}, a^i=0.\label{ckappa}
\end{equation}
From now onwards, we {\it do~not} write``hat" explicitly over the coordinates of the $\kappa$-spacetime since there is no scope of confusion as all the coordinates we use are that of noncommutative spacetime. It can be easily seen that the conditions imposed on $c^{\mu\nu}_{~~\lambda}$ by the Jacobi identity involving only coordinates are trivially satisfied.

Differentiating the Eqn.(\ref{kappacoord}) with respect to $\tau$, we get $[{\dot x}^\mu, x^\nu] +[ x^\mu, {\dot x}^\nu]= i c^{\mu\nu}_{~~\lambda}{\dot x}^\lambda$.For $c^{\mu\nu}_{~~\lambda}=0$, the above equation is symmetric in $\mu$ and $\nu$. For the present case we get,
\begin{equation}
[{\dot x}^\mu,  x^\nu]=-\frac{i \hbar}{m_{eff}}g^{\mu\nu}+ i c^{\mu\nu}_{~~\lambda}{\dot x}^\lambda\label{kfb},
\end{equation}
where $m_{eff}=m(1+amc)$( see discussion below). Note that the functional form of $g^{\mu\nu}$ is not fixed by the differentiation of Eqn.(\ref{kappacoord}), but fixes only the anti-symmetric part. In the above, $g^{\mu\nu}$ can, in general, be a function of both coordinates as well as velocities. Now, we impose the  requirement that the above equation reproduces the well known result in the commutative limit. This imply that  as $a\to 0$, $g^{\mu\nu}\to \eta^{\mu\nu}$. The Jacobi identity involving a velocity and two coordinates show that the $g^{\mu\nu}$ can be just $\eta^{\mu\nu}$ itself. Though in the original approach, momentum was not assumed to exist, here, we take the known commutation relation between coordinate and momenta \cite{kbasis} as a guiding tool to fix the dependence of $g^{\mu\nu}$ on $x$ and ${\dot x}$. We also note that the non-relativistic limit of the energy-momentum relation in the $\kappa$-spacetime  shows that the momentum, up to first order in $a$, is $p_i=m_{eff}{\dot x}_i$, where $m_{eff}=m(1+amc)$\cite{akkeh}. This suggest that up to first order in $a$, $g^{\mu\nu}$ is just equal to $\eta^{\mu\nu}$ in the above equation. This also shows that the higher order (in $a$) corrections to  $g^{\mu\nu}$ depend only on velocities and not on coordinates themselves. With these considerations, we take the Feynman brackets, valid up to first order in the deformation parameter $a$ to be
\begin{equation}
[x^\mu, {\dot x}^\nu]=\frac{i \hbar}{m_{eff}}\eta^{\mu\nu}+ i c^{\mu\nu}_{~~\lambda}{\dot x}^\lambda\label{kfb1}.
\end{equation}
Differentiating the above equation with respect to $\tau$ gives
\begin{equation}
[x^\mu, {\ddot x}^\nu]=-[{\dot x}^\mu, {\dot x}^\nu] +i c^{\mu\nu}_{~~\lambda}{\ddot x}^\lambda.\label{kfs1}
\end{equation}
The right side of the above equation is anti-symmetric in indices and the above equation allow us to define a rank two, anti-symmetric tensor as in the commutative case. Thus we define
\begin{equation}
[{\dot x}^\mu, {\dot x}^\nu] =\frac{i\hbar q}{m_{eff}^2} F^{\mu\nu} + i c^{\mu\nu}_{~~\lambda}{\ddot x}^\lambda.\label{kfs}
\end{equation}
Using Eqn.(\ref{kfs}) in Eqn.(\ref{kfs1}) and integrating leads to 
\begin{equation}
F(x,{\dot x})^\mu=G(x)^\mu+q F^{\mu}_{~\nu}{\dot x}^\nu\label{kne}
\end{equation}
where we have assumed the Newtons second law, valid up to first order in $a$ to be $F^\mu=m_{eff}{\ddot x}^\mu$. In the above, $F_{\mu\nu}$ can be function of coordinates and velocities. Using the Jacobi identity $[x^\mu,[{\dot x}^\nu,{\dot x}^\lambda]]+$ cyclic terms, shows that $[x^\mu, F^{\nu\lambda}]\ne 0$. In the limit $a\to 0$, one gets $[x^\mu, F^{\nu\lambda}]= 0$, implying $F^{\mu\nu}$ is  a function of coordinates alone, in the commutative spacetime. Thus, in the present case, $a$ dependent terms in $F^{\mu\nu}$ can be functions of $x$ as well as ${\dot x}$. We take $F^{\mu\nu}$ as function of $x$ alone, as a first approximation, in deriving Eqns. (\ref{khme}) and (\ref{kinhme}). Had we taken $F^{\mu\nu}$ to be function of ${\dot x}$ also, we would have got additional terms of the form $(i\hbar/ m_{eff}^2)\frac{\partial F^{\mu\nu}}{\partial{\dot x}^\mu} +i\hbar c^{\mu\alpha}_{~~\rho}{\dot x}^\rho\frac{\partial F^{\nu\lambda}}{\partial {\dot x}^\alpha}$. Here the second term is quadratic in $a$ and hence we neglect it. The first term has an additional $\hbar/m_{eff}$ compared to remaining terms of Eqns. (\ref{khme}) and (\ref{kinhme}) and thus  negligible. This justify taking $F^{\mu\nu}$ as a function of $x$ alone.

To derive the Maxwell's equations, we use the Jacobi identity involving velocities alone, i.e., 
\begin{equation}
[{\dot x}^\mu,[{\dot x}^\nu, {\dot x}^\lambda]]+[{\dot x}^\nu,[{\dot x}^\lambda, {\dot x}^\mu]]+[{\dot x}^\lambda,[{\dot x}^\mu, {\dot x}^\nu]]=0.
\end{equation}
Using Eqn.(\ref{kfs}) in the above equation gives the $\kappa$-deformed, homogeneous Maxwell's equations (up to first order in deformation parameter) as
\begin{eqnarray} 
\frac{\hbar}{m_{eff}}(\partial^\mu F^{\nu\lambda}+\partial^\nu F^{\lambda\mu}+\partial^\lambda F^{\mu\nu})&=&
\left[ c^{\mu\alpha}_{~~\rho}{\dot x}^\rho\partial_\alpha F^{\nu\lambda}+
c^{\nu\alpha}_{~~\rho}{\dot x}^\rho\partial_\alpha F^{\lambda\mu}
+c^{\lambda\alpha}_{~~\rho}{\dot x}^\rho \partial_\alpha F^{\mu\nu}\right]\nonumber\\
&-&\frac{im_{eff}^2}{\hbar q}\left[ c^{\nu\lambda}_{~~\rho}[{\dot x}^\mu, {\ddot x}^\rho]+ c^{\lambda\mu}_{~~\rho}[{\dot x}^\nu,{\ddot x}^{\rho}]+ c^{\mu\nu}_{~~\rho}[{\dot x}^\lambda,{\ddot x}^\rho]
\right].\label{khme}
\end{eqnarray}
 Contracting $\mu$ with $\nu$ by multiplying above equation with $\eta_{\mu\nu}$ gives
  \begin{eqnarray}
\partial_\mu F^{\mu\lambda}-\frac{m_{eff}}{\hbar}c^{\mu\alpha}_{~~\rho}
{\dot x}^\rho\partial_\alpha F_{\mu}^{~\lambda} +\frac{im_{eff}^3}{q\hbar^2} c^{\nu\lambda\rho} [{\dot x}_\nu, {\ddot x}_\rho]
+\partial_\mu F^{\lambda\mu}-\frac{m_{eff}}{\hbar}c^{\mu\alpha}_{~~\rho}{\dot x}^\rho\partial_\alpha F^{\lambda}_{~\mu} +\frac{im_{eff}^3}{q\hbar^2}c^{\lambda\nu\rho}[{\dot x}_\nu, {\ddot x}_\rho]=0.\label{kappacur}
\end{eqnarray}
Using anti-symmetry of indices, we see that the left hand side of the above equation vanishes identically. We use the above equation to {\it define} the current\cite{jmp40}, thereby obtaining the inhomogeneous equations, i.e.,
\begin{equation}
[\partial_\mu F^{\mu\nu}-\frac{m_{eff}}{\hbar}c^{\mu\alpha}_{~~\rho}{\dot x}^\rho\partial_\alpha F_{\mu}^{~~\nu}]=
-[\partial_\mu F^{\nu\mu}-\frac{m_{eff}}{\hbar}c^{\mu\alpha}_{~~\rho}{\dot x}^\rho\partial_\alpha F^{\nu}_{~~\mu}]\doteq
-j^\nu.\label{kinhme}
\end{equation}
In obtaining the above equation from Eqn.(\ref{kappacur}), we have used the anti-symmetry in the first  two indices in $c^{\mu\nu}_{~~\lambda}$ to cancel terms involving $[{\dot x}, {\ddot x}]$.  This complete the derivation of  both homogeneous and inhomogeneous Maxwell's equation on $\kappa$-spacetime, in presence of source and a moving particle of charge $q$ and mass $m$. Note these equations are  valid up to first order in the deformation parameter. Also note that in the limit $a\to 0$, we recover the known equations in the commutative space.

We now turn our attention to Newtons equation of motion. Using Eqn.(\ref{kne}) and the fact that $F^\mu=m_{eff}{\ddot x}^\mu$, we find
\begin{eqnarray}
[{\dot x}^\mu, G^\nu]-[{\dot x}^\nu, G^\mu]&=&\frac{i\hbar}{m_{eff}}(\partial^\mu F^{\nu\lambda}+{\rm cyclic~in~}\mu,\nu,\lambda){\dot x}_\lambda 
+{i\hbar q}\left[ c^{\mu\alpha}_{~~\rho}{\dot x}^\rho\partial_\alpha F^{\nu\lambda}{\dot x}_\lambda -\mu \leftrightarrow \nu\right]\nonumber\\
&-&i\hbar q\left[ F^{\nu\lambda}c^{\mu}_{~\lambda\alpha}{\ddot x}^\alpha -\mu\leftrightarrow \nu\right]\label{ge1}
\end{eqnarray}
Since the $G^\mu(x)$ in Eqn.(\ref{kne}) is a function of $x$ alone, we find
\begin{equation}
[{\dot x}^\mu, G^\nu]=-\frac{ih}{m_{eff}}\left(\partial^\mu G^\nu+\frac{1}{\hbar}c^{\mu\alpha}_{~~\lambda}{\dot x}^\lambda\partial_\alpha G^\nu\right).\label{ge2}
\end{equation}
Using above equation in Eqn.(\ref{ge1}),  we find,
\begin{equation}
\partial_\mu G_\nu-\partial_\nu G_\mu = O(a)
\end{equation}
showing that, once the first order correction in the deformation parameter is included, $G^\mu=\partial^\mu\Phi+ O(a)$  where $\Phi$ is a scalar. This shows that the interaction of quantum mechanical particle in $\kappa$-spacetime with the scalar field is modified. We also note that the  Eqn.(\ref{khme}) show that $F_{\mu\nu}=\partial_\mu A_\nu-\partial_\nu A_\mu+ O(a)$, showing that the gauge field interaction is also modified in the $\kappa$-spacetime.

\section{Electric-Magnetic Duality}

In this section, we analyze the $\kappa$-deformed Maxwell's equations obtained in Eqns.(\ref{khme},\ref{kinhme}).  Using explicit form of $c^{\mu\nu}_{~~\lambda}$ given in Eqn.(\ref{ckappa}), in Eqns. (\ref{khme},\ref{kinhme}), we get the Maxwell's equations, valid up to {\it first order} in the deformation parameter as
\begin{eqnarray}
{\vec \nabla}\cdot {\vec B}+\frac{m a}{\hbar}{\vec v}\cdot\partial_0{\vec B}=0,\label{kg1}\\
\partial_0{\vec B}+ {\vec\nabla}\times{\vec E}+\frac{m a}{\hbar}\left[ v^{i}\partial_i{\vec B}
+{\vec v}\times\partial_0{\vec E}\right]=0\label{kf}\\
{\vec \nabla}\cdot {\vec E} +\frac{m a}{\hbar}{\vec v}\cdot\partial_0{\vec E}=\rho_{e}\label{kg2}\\
\partial_0{\vec E}-{\vec\nabla}\times{\vec  B}+\frac{m a}{\hbar}\left[ v^i\partial_i{\vec E} -{\vec v}\times\partial_0{\vec B}\right]=-{\vec j}_{e}\label{kappameqn}
\end{eqnarray}
Here, we have used ${x}^i/d\tau=v^i$. We note that in the limit $a\to 0$, above equation reproduce the well known  Maxwell's equations in the  commutative spacetime. It is interesting to see that the mass of the charged particle also appears naturally in the $\kappa$-dependent of the modified Maxwell's equation. This suggest that the laws of electrodynamics for particles with same charge, but with different masses will be different. This result is a complete departure from the commutative spacetime\footnote{We note that, by
re-expressing the above, modified Maxwell's equations in terms of $LT^{-1}u^i= v^i$, the new terms of the deformed Maxwell's equations can be parameterized by a dimensionless quantity ${\tilde a}=(ma/\hbar) LT^{-1}$.}. Also, it is interesting to note that for the static electric and magnetic fields, the Gauss law equations (\ref{kg1} and \ref{kg2}) do not get any $a$ dependent modifications while Eqns. (\ref{kf}, \ref{kappameqn}) are still modified due to $\kappa$-deformation.

In the commutative space, Maxwell's equation, in the absence of source is invariant under the combined transformations ${\vec E}\to {\vec B}, {\vec B}\to{-\vec E}$. This is a symmetry even when sources are present,  provided that the magnetic monopole exist  and  along with above transformation,  we also have  $\rho_{e}, j_{e}\to \rho_{mag}, j_{mag};  \rho_{mag}, j_{mag} \to-\rho_{e}, -j_{e}$.  It is easily seen from the above Eqns. (\ref{kg1},\ref{kf}, \ref{kg2}) and (\ref{kappameqn}) that the $a(=\kappa^{-1})$ dependent terms maintain this symmetry.

\section{Lorentz Symmetry}
It is well known \cite{plaijtpa} that the non-vanishing brackets between velocities  given in Eqn.(\ref{fs}), modifies the Lorentz algebra
\begin{eqnarray}
[L^{\mu\nu}, L^{\lambda\sigma}]=i\hbar \left (\eta^{\mu\sigma} L^{\lambda\nu}-\eta^{\mu\lambda}L^{\sigma\nu} +\eta^{\nu\lambda}L^{\sigma\mu}-
\eta^{\nu\sigma}L^{\lambda\mu}\right)
+i\hbar q\left( x^\mu x^\lambda F^{\nu\sigma}-x^\nu x^\lambda F^{\mu\sigma}-x^\mu x^\sigma F^{\nu\lambda}+x^\nu x^\sigma F^{\mu\lambda}\right),\label{ll}
\end{eqnarray}
\begin{equation}
[x^\mu, L^{\nu\lambda}]=i\hbar\left(x^\nu\eta^{\mu\lambda}-x^\lambda\eta^{\mu\nu}\right),\label{xl}
\end{equation}
\begin{equation} 
[{\dot x}^\mu, L^{\nu\lambda}]=-i\hbar \left(\eta^{\mu\nu}{\dot x}^\lambda-\eta^{\mu\lambda}{\dot x}^\nu\right) +\frac{i\hbar}{m}\left( x^\nu F^{\mu\lambda}-x^\lambda F^{\mu\nu}\right)
\label{vl}
\end{equation}
in the commutative spacetime itself.  In this section we study the consequences of $\kappa$-deformation on the above algebra. As we have seen in the previous sections, the $\kappa$-deformation drastically alter the Maxwell's equations as well as the force equation. Thus, it is of interest to see how $\kappa$-deformation affect the above algebraic structure.

It is clear that the $\kappa$-dependent modifications to the commutators in Eqns.(\ref{kappacoord},\ref{kfb1},\ref{kfs}) do modify the above algebra in the $\kappa$-spacetime, and can be calculated in by a straightforward, but lengthy calculation.

Now, we modify the generators of the algebra to be
\begin{equation}
{\cal L}^{\mu\nu}=L^{\mu\nu}+ M^{\mu\nu}\label{modifiedgen}
\end{equation}
such that ${\cal L}^{\mu\nu}$ will satisfy same commutation relations as in the commutative spacetime given in Eqns.({\ref{ll},\ref{xl},\ref{vl})(such modifications of 
symmetry algebra have been studied with different motivations in \cite{wess, sm, sb}). With this requirement, we find
\begin{eqnarray}
[x^\mu, M^{\nu\lambda}]&=&-im_{eff}\left[ c^{\mu\nu}_{~~\alpha} x^\alpha{\dot x}^\lambda-c^{\mu\lambda}_{~~\alpha} x^\alpha{\dot x}^\nu\right]
+i\hbar m_{eff}\left[ c^{\mu\lambda}_{~~\alpha}x^\nu x^\alpha -c^{\mu\nu}_{~~\alpha}x^\lambda x^\alpha\right]\nonumber\\
&=& i\hbar(m^{-1}-m_{eff}^{-1})\left(x^\nu\eta^{\mu\lambda}-x^\lambda\eta^{\mu\nu}\right)\label{xm}
\end{eqnarray}
\begin{eqnarray}
[{\dot x}^\mu, M^{\nu\lambda}]&=&-i\frac{\hbar}{m_{eff}}q \left[x^\nu F^{\mu\lambda}-x^\lambda F^{\mu\nu}\right]-i\hbar m_{eff} x^\alpha \left[c^{\mu\nu}_{~~\alpha} {\dot x}^\lambda-
c^{\mu\lambda}_{~~\alpha}{\dot x}^\nu\right]\nonumber\\
&+&i\hbar m_{eff}\left[c^{\mu\lambda}_{~~\alpha} x^\nu-c^{\mu\nu}_{~~\alpha} x^\lambda\right]{\ddot x}^\alpha\label{dotxm}
\end{eqnarray}
Above two equations along with the condition that ${\cal L}^{\mu\nu}$ should have same commutation relations as in Eqn.({\ref{ll}) gives
\begin{equation}
 M_{jk}= q\left[x^ix_jF_{ik}-x_i x^i F_{jk}+x^i x_k F_{ji}\right]\label{magang}
\end{equation}
where we have taken, up to first order in deformation parameter, $[M_{ji}, M_{ik}]=0$, which is true since $[x_i, x_j]=0$ in the $\kappa$-spacetime. Thus we find
\begin{equation}
M_i=\epsilon_{ijk}M^{jk}=-q (x\cdot B) x_i\label{mag}
\end{equation}
where $\frac{1}{2}\epsilon_{ijk}F^{jk}=B_i$. Consistency of $[{\dot x}^i, M^j]$ calculated using the explicit form of $M^j$ given above with the same derived from Eqn.(\ref{dotxm}) leads to the condition on the magnetic field as
\begin{equation}
x^iB_m+x_mB^i=-x^j\partial_i B_j x_m.
\end{equation}
This is satisfied by 
${\vec B}=g\frac{\vec r}{4\pi r^3}.$
Using this in Eqn.(\ref{mag}), gives the Poincare magnetic angular momentum\cite{plaijtpa}
${\vec M}=-qg\frac{\vec r}{4\pi r^2}.$
Here we note that magnetic angular momentum do not get any modification due to $\kappa$-deformation. This is expected since the space coordinates have vanishing commutators among themselves in $\kappa$-spacetime.  From the fact that $a^0=a$ , it is easy to see from Eqns.(\ref{xm}, \ref{dotxm}) that the boost sector will get modification due to the $\kappa$-deformation, but these modifications may not have interesting interpretation unlike that of angular momentum sector, since even in the commutative spacetime, the boost sector of $M^{\mu\nu}$ in Eqn.(\ref{modifiedgen}), do not have any interesting interpretation\cite{plaijtpa}.

\section{Conclusions}

We have derived the Maxwell's equations on $\kappa$-Minkowski  spacetime, using the Feynman's approach. We have also obtained the force equation
for particle moving in this spacetime.  These results are valid up to first order in the deformation parameter. We have also analysed the magnetic momentum generator needed to maintain the Lorentz algebra relations in the $\kappa$-spacetime. 

The $\kappa$-deformed Maxwell's equations shows that the electric-magnetic duality symmetry is present in the $\kappa$-spacetime also(up to first order in deformation parameter). But for the specific case where the electric and magnetic fields are time independent, this symmetry is broken in $\kappa$ spacetime, unlike in the commutative spacetime.  In the Moyal spacetime, electric-magnetic duality had been studied and  the dual of $U(1)$ gauge theory was constructed \cite{OJGRabin}.

The $\kappa$-deformed Maxwell's equations, further show,  the laws of electrodynamics will be different for particles with same electrical charges but having different masses. This effect of $\kappa$-deformation on electrodynamics is distinctly different from the modifications due to Moyal spacetime. Certainly this feature is drastically different from the commutative spacetime and hence can lead to stringent limits\cite{bounds} on the deformation parameter(studies along these lines are in progress). Modified force equation (Eqn.(\ref{kne})) shows that the particle's interactions with scalar and gauge fields get modified due to the $\kappa$-deformation.

\noindent {\bf Acknowledgment}: EH thank A. Khare and M. Sivakumar for useful comments. EH also thank the anonymous referee for useful comments and suggestions.

\end{document}